\documentclass[conference]{IEEEtran}
\usepackage[cmex10]{amsmath}
\usepackage{amssymb}
\usepackage{graphicx}
\usepackage{hyperref}
\usepackage{balance}

\newcommand{\remove}[1]{}

\newcommand{\fig}[1]{Fig.~\ref{#1}}
\newcommand{\eq}[1]{Eq.~\ref{#1}}
\newcommand{\tbl}[1]{Table~\ref{#1}}

\newcommand{\figwidth}{0.9\columnwidth} 
\newcommand{\Binomial}{\mbox{Bi}} 
\newcommand{\Rreceived}{R_{\mbox{\scriptsize posts received}}}	
\newcommand{\Rposts}{R_{\mbox{\scriptsize posts}}}	
\newcommand{\Rpostsbar}{\bar{R}_{\mbox{\scriptsize posts}}}	
\newcommand{\Rvisits}{R_{\mbox{\scriptsize visits}}}	
\newcommand{\Nfriends}{N_{\mbox{\scriptsize friends}}}	
\newcommand{\ViewsPerPost}{\nu}				
\newcommand{\Prob}{\mbox{Prob}}	
\newcommand{\Ptopic}{P_{\mbox{\scriptsize topic}}}	
\newcommand{\Pact}{P_{\mbox{\scriptsize act}}}	

\newcommand{\Pposts}{P_{\mbox{\scriptsize posts}}}	

\newcommand{\Pview}{P_{\mbox{\scriptsize view}}}	
\newcommand{\Pvisible}{P_{\mbox{\scriptsize visible}}}	
\newcommand{\Pinteresting}{P_{\mbox{\scriptsize interesting}}}	
\newcommand{\Prespond}{P_{\mbox{\scriptsize respond}}}	

\begin{document}
\title{Stochastic Models Predict User Behavior\\ in Social Media}
\author{
\IEEEauthorblockN{Tad Hogg}
\IEEEauthorblockA{Institute for Molecular Manufacturing\\
Palo Alto, CA\\
tadhogg@yahoo.com}
\and
\IEEEauthorblockN{Kristina Lerman}
\IEEEauthorblockA{Information Sciences Institute\\
Marina Del Rey, CA 90292\\
lerman@isi.edu}
\and
\IEEEauthorblockN{Laura M. Smith}
\IEEEauthorblockA{Information Sciences Institute\\
Marina Del Rey, CA 90292\\
lausmith@fullerton.edu}
}

\maketitle
\begin{abstract}
User response to contributed content in online social media depends on many factors. These include how the site lays out new content, how frequently the user visits the site, how many friends the user follows, how active these friends are, as well as how interesting or useful the content is to the user. We present a stochastic modeling framework that relates a user's behavior to details of the site's user interface and user activity and describe a procedure for estimating model parameters from available data.
We apply the model to study discussions of controversial topics on Twitter, specifically, to predict how followers of an advocate for a topic respond to the advocate's posts. We show that a model of user behavior that explicitly accounts for a user transitioning through a series of states before responding to an advocate's post better predicts response than models that fail to take these states into account. We demonstrate other benefits of stochastic models, such as their ability to identify users who are highly interested in advocate's posts.

\end{abstract}


\section{Introduction}
\remove{
\note{some ideas for introduction:}

Applying models of user behavior in social media: especially relevant but challenging for short-term group activities where there is limited time for collecting detailed data to average over, there is large user turnover and variation in activity level related to external events (e.g., a campaign).

Can models that require estimating parameters give useful insight in behavior in such rapidly changing social media situations? How to estimate parameters from available data? Instead of detailed prediction for response to individual content as in prior studies, here we focus on average behavior of the community response.
}

More data about social behavior is now available than ever before. These data, much of which come from social media sites such as Twitter, contain traces of individual activity and social interactions. On Twitter, interactions include users posting short text messages, called tweets, and  following other users to receive their posts. Users  may also respond to posts shared by others, for example, by retweeting them to their own followers. These abundant data offer new opportunities for learning models of user behavior and interests, identifying communities of like-minded users, and inferring the topics of conversations between them. The models can, in turn, be used to understand how popular opinion is changing, predict future activity, and identify timely and interesting information.

Researchers have developed a variety of probabilistic methods to learn user models from social data~\cite{Lauw12,Kang13sbp,WangB11,purushotham}. Such models usually include a user's interest in some topic as a hidden parameter, which is estimated from her response to messages on that topic. The more a user responds, for example, by retweeting a message on the topic, the more she is interested in it. These models, however, fail to account for details of user behavior that affect response, such as how often the user visits Twitter, how many messages she receives, and how many of these she inspects. Without considering these variables, it is difficult to explain behavior. Does a lack of response mean that a user is not interested in the topic, or that she simply did not see the message? These probabilistic models also need large amounts of data to learn accurate models, which may not be obtainable for all users.

Stochastic modeling is an alternative approach to modeling user behavior. It is a probabilistic framework that represents each user as a stochastic process that transitions between states with some probability. The probabilistic representation captures our uncertainty about individual actions. When modeling a social media site, the user's states, and therefore, behavioral outcomes, are constrained by the user interface.   On Twitter, the states include visiting  the site, seeing a post, and responding to it, e.g., by retweeting it. Transitions represent dependencies between states: e.g., responding to a post is conditioned on seeing it and being interested in it.

The model contains parameters that govern, for example, how frequently the user visits Twitter. These parameters capture the salient mechanisms driving behavior: e.g., seeing  a post depends on its position in the user's feed and the likelihood the user will navigate to that position, which depends on how deeply users explore their feeds.  Model parameters could be user-specific, but to reduce the required data and improve generalization, we associate parameters with populations of similarly-behaving users.

We estimate parameters from data using maximum likelihood. One challenge of this approach is lack of available data for parameter estimations. For example, we do not know when a user visits Twitter, only when she posts or retweets a message, and must, therefore, estimate visit rate from these data. However, we show that even such crude estimates lead to useful models of user behavior.

In the past, we used a stochastic modeling framework to describe dynamics of popularity of content on social media~\cite{hogg12b,Lerman12tist}. In this paper we adapt the model to describe individual user behavior. We describe how we estimate parameters of the model and use the model to predict how users will respond to posts about a specific topic. We demonstrate that a model that accounts for the likelihood of seeing posts and user's interest better predicts response than just using the user's activity.

While in this paper we only consider predicting response, there are many other interesting applications of the stochastic modeling framework. For example, this approach could help identify dedicated users and suggest strategies that would amplify user response.

\section{Data}
\label{sec.data}

Our data consist of Twitter posts related to initiatives that appeared on the November 2012 California ballot. A ballot initiative, or \emph{proposition}, is a political process that allows citizens of some states, including California, to place new legislation on the ballot. If the proposition wins the popular vote, it becomes law. A total of eleven propositions appeared on the 2012 California ballot, including two that would raise taxes to fund public schools (Propositions 30 and 38), one that would require manufacturers to label products that contain genetically modified organisms (GMOs) (Proposition 37), a proposition to abolish the death penalty (Proposition 34), and one to repeal the three strikes law (Proposition 36), among others. This domain is a convenient choice for our study: discussions about topics (propositions) mainly occur over a limited time period immediately preceding the election when there is concentrated interest in the topic, each proposition has a clearly identified advocate, representing either pro and con positions, with a presumed intention to influence followers' behavior, e.g., in spreading the advocate's message.

We began to collect data in August 2012, using the following strategy. First, we created terms related to proposition names, including permutations of the terms ``proposition'' or ``prop'', numeric value, and ``yes'' or ``no'' to indicate stance: e.g., `prop30'', ``proposition37'', ``yeson30'', ``noprop32''. Next, we identified topical terms related to propositions such as ``ca2012'' and ``nonewtaxes''. We then extended the set of terms to include those that occur frequently with these terms. This strategy identified additional relevant terms such as ``righttoknow'', ``LabelGMOs'', ``voteyeson37'', and ``stopspecialexemptions'', for a total of 95 terms. Using the Search API, we monitored Twitter to collect tweets that contained these terms as hashtags or keywords.

\begin{figure}
\centering
  \includegraphics[width=\figwidth]{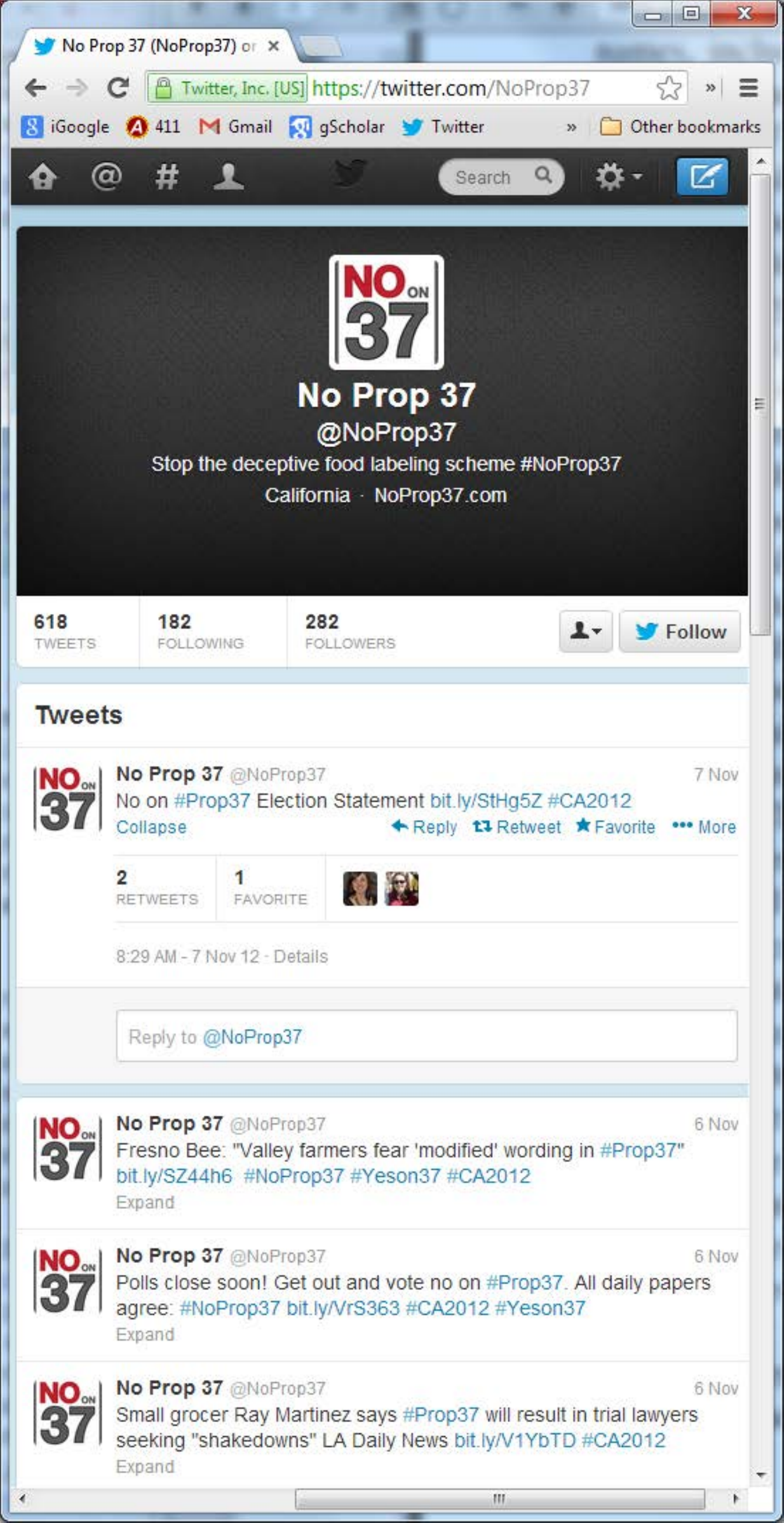}
\caption{Twitter account for ``@NoProp37'', retrieved August 12, 2013. \label{fig.screenshot}}
\end{figure}

In addition to monitoring Twitter feed for mentions of topics related to the propositions, we also identified campaign advocates, that is Twitter accounts that explicitly promoted a position on the topic, and added them to the list of monitored accounts. Some of the advocates were easy to identify, such as ``@YesOnProp30'' or ``@NoProp37''.  We found others by reviewing similar account suggestions made by Twitter, e.g., ``@CARightToKnow''. We then manually reviewed these accounts to verify their position on the issue, and added them to monitored accounts. Figure~\ref{fig.screenshot} shows the Twitter account for the advocate against Proposition 37. Note that information about this advocate was retrieved in August, 2013. The time delay accounts for the larger observed number of followers.

We collected 44M tweets made by 600K monitored accounts, creating a complete record of their activity in the months preceding the November election.  In addition, we periodically retrieved the names of followers of the 81 advocate accounts.  We collected all tweets produced by these followers.

\remove{
\begin{table}
\centering
\begin{tabular}{|c|c|c|c|}
\hline
users &	tweets &	advocates &	followers \\ \hline
599,629	& 43,751,870 &	81	 & 899,888 \\ \hline
\end{tabular}
\caption{Data collected from Twitter}\label{tbl:all-data}
\end{table}
}

\begin{table}[!htb]
\centering
\begin{tabular}{@{}l@{}ccc@{}}
\hline
	& {\it number } & {\it number } & {\it typical activity}\\
{\it advocate}  &{\it of followers} & {\it of posts}   & {\it rate of followers}\\ \hline\\
@YesOnProp30 & 108	& 422 & 1.61   \\
\\
 \multicolumn{4}{l}{\quad``Pass $\#$Prop30 to invest in me and the future of my generation of } \\
\multicolumn{4}{l}{\quad Californians! http://t.co/5fp2WSRb'' }\\
\multicolumn{4}{l}{\quad {\it Number of RTs}  32}\\
\hline \\
@StopProp30 & 42 & 38 & 1.51 \\
\\
 \multicolumn{4}{l}{\quad ``Fiscal Feud! Prop 30 Politicians Prove We'll Never Know Where}\\
 \multicolumn{4}{l}{\quad  the Money Goes. Vote NO on $\#$Prop30 http://t.co/hQboqP5G''}\\
 \multicolumn{4}{l}{ \quad{\it Number of RTs}  7}\\
 \hline \\
 @CARightToKnow  & 215  	& 1520 & 8.32 \\
 \\
  \multicolumn{4}{l}{\quad  ``Russia Suspends Use of GMO corn made by $\#$Monsanto  due to }\\
   \multicolumn{4}{l}{\quad  study linking GMOs to cancer.  http://t.co/Ob9y... RT! $\#$YesOn37 }\\
  \multicolumn{4}{l}{\quad   $\#$LabelGMOs''}\\ 
 \multicolumn{4}{l}{ \quad{\it Number of RTs}  65}\\
 \hline \\

@NoProp37 & 94		& 391 & 1.78
\\
\\
  \multicolumn{4}{l}{\quad  ```Even voters who worry about genetically modified food should reject'}\\
  \multicolumn{4}{l}{\quad $\#$Prop37. Sac Bee editorial http://t.co/...  $\#$Yeson37 $\#$NoProp37''}\\
   \multicolumn{4}{l}{ \quad{\it Number of RTs}  8}\\

\hline
\end{tabular}
\caption{Features of advocate accounts, including the number of followers they have, the number of messages posted during the data collection period, and examples posts. These examples are among the most retweeted posts by each advocate.  The number below each example post is the number of times this particular tweet was retweeted by other users.}\label{tbl.data}
\end{table}

\remove{
\begin{table}
\centering
\begin{tabular}{lcc}
\hline \\
	& \multicolumn{2}{c}{\emph{number of}} \\
\emph{advocate}	 		& \emph{followers} & \emph{advocate posts} \\ \hline
YesOnProp30		& 108	& 422 \\
StopProp30		&  42		& 38 \\
CARightToKnow	& 215  	& 1520\\
NoProp37			& 94		& 391\\
\end{tabular}
\caption{Data sets.}\label{tbl.data}
\end{table}
}

This paper studies a subset of these data: specifically, select advocates linked to the most actively discussed propositions and examines the response of their followers to their posts.
Table~\ref{tbl.data} gives the names of the selected advocate accounts, the number of posts made by them, and the number of their followers we studied.  Each of these followers  had been following the respective advocate since mid-September and had posted on the proposition, enabling us to determine the user's stance (pro, con, neutral) on the proposition manually from their tweets. Neutral users are typically journalists, who often follow advocates holding different stances, allowing them to observe the opposing arguments.

Table~\ref{tbl.data} also provides a sample tweet for each advocate, demonstrating the different agendas of the advocates and their follower responses.    These tweets are selected from the set of advocates' tweets with the most retweets. Note that advocate tweets, as well as tweets from followers, often contain terms and hashtags related to the proposition. We use the presence of these keywords to determine the level of user's interest in the topic of the proposition. The ease of measuring such targeted interests is an advantage of examining discussions of propositions, compared to attempting to determine the topic content of general tweets.

In addition to classifying whether tweets relate to a given proposition, we need to identify the followers' response to received posts. We did so by focusing on posts from the advocate and determining whether the follower retweeted each one. We did so by examining each post from each follower to identify which posts were retweets of the advocate's post and the original advocate post.  Retweets typically have the structure ``RT @username: retweeted text,'' where @username is the original posting user.  This ``RT'' may be replaced by other variations such as ``Retweet'', ``Retweeting'', and ``via.'' Searching for this structure identified the follower responses.

\section{Stochastic Modeling Framework}

The behavior of individual users on social media depends on their history of interactions with other users and content provided by the site, and varies considerably over the user population~\cite{joyce06,wilkinson08}. Stochastic models summarize this behavior by representing an individual entity, whether a user or contributed content, as a stochastic process with a few states~\cite{Lerman07ic,hogg09c,Iribarren09,social-physics,hogg12b}. Usually these models have the Markov property, where the future state of a user or contributed content depends only on the present state and the input from the site and other users at that time. A Markov process is succinctly captured by a \emph{state diagram} showing the possible states and transitions between those states.
This approach is similar to compartmental models in biology~\cite{ellner06} and population dynamics~\cite{social-physics}. E.g., in epidemiology such models track the progress of a disease as individuals transition between states, such as susceptible and infected.

Developing a model for social media requires identifying the key states and relevant transitions between them. The states are related to the design of the social media site, since its interface constrains users's actions.
The state diagram has a probabilistic interpretation. Each state represents the probability of a user performing that action, with transitions leading to conditional probabilities. The probabilistic interpretation of the model of user behavior allows us to compute the likelihood the user performed a given observed action, e.g., responded to a post, and then estimate model parameters using maximum likelihood. The remainder of this section describes our model for evaluating how followers respond to an advocate's posts, and how we estimate the transition rates from our data.

\subsection{Model}

\begin{figure}
\centering
  \includegraphics[width=\figwidth]{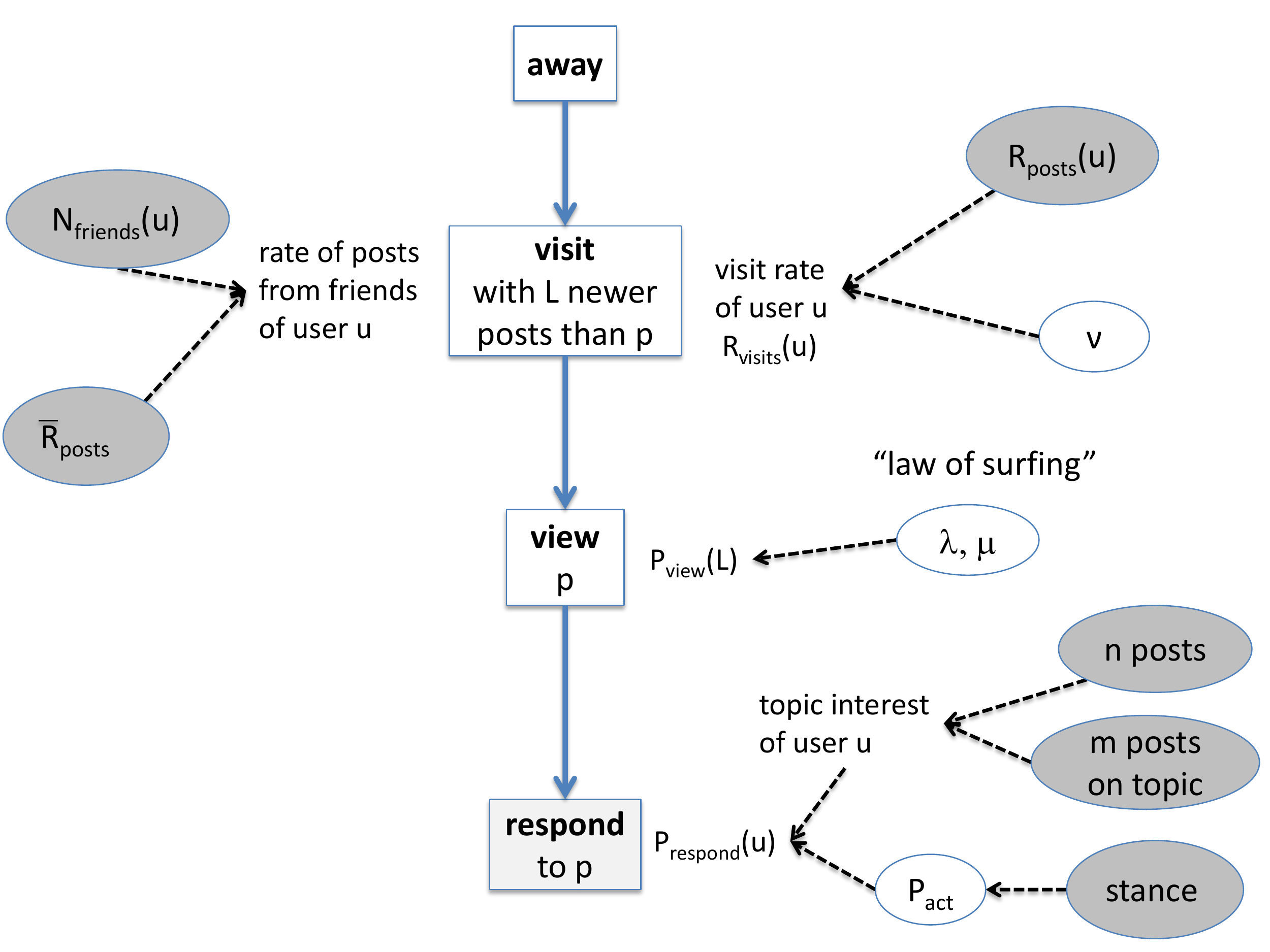}
\caption{State diagram for model of user behavior with respect to a single post received from a friend. Boxes represent states a user $u$ transitions through before making an observable action, such as respond to a post $p$. Ovals represent parameters of the model, some of which are measured (grey) and others estimated (white) from data. \label{fig.state diagram}}
\end{figure}

In this study, we focus on {\it user} behavior, in contrast to prior research with stochastic models that focused on the collective response of social media users to shared content.  \fig{fig.state diagram} shows the steps through which a user $u$ from a {\it  population of followers} of some advocate $a$ progresses to handle newly received content, e.g., a new post $p$ from $a$. The state diagram captures the salient details of the site's user interface and how $u$ interacts with it. When the advocate posts a new message $p$, the web site adds it to $u$'s list of new content, making it {available} for viewing. By the time of $u$'s later \textbf{visit}, $u$'s {\it friends}, i.e., other accounts $u$ follows, will have generated some number, $L$, of newer posts, moving $p$ to the $L+1^{st}$ position in user $u$'s list. The user may examine enough of this list to \textbf{view} $p$. Once viewed, the user may decide to \textbf{respond} to it, e.g., by forwarding it to followers via a retweet.

User {activity} in social media, i.e., the rate  they post content on the site, varies systematically over the course of a day as well as on longer time scales. One way to simplify models by making activity rates more uniform is to redefine the unit of time to correspond to overall activity on the site~\cite{szabo09}. For our study, we average over several months of activity so these short term variations are not important. Moreover, the analysis only involves {relative} rates of users' activity and their friends, so this time adjustment would not change the analysis. This focus on relative rates also adjusts for systematic increases in activity as the polling date approached.

Behavior varies considerably across users and content. Thus an important modeling choice is how much of this variation to include explicitly while averaging over the rest. For instance, in communities where content is readily visible to all users, not just those who follow the submitter, a useful grouping of users is whether they are followers of the content's submitter~\cite{hogg12b}. For Twitter, a major distinction is the number of friends a user has~\cite{Hodas12socialcom}, since this strongly affects the rate at which the user receives posts (\eq{eq.received}).

For this paper, we compartmentalize, or group, users by their \emph{activity}, \emph{interest} in the topic and \emph{number of friends}. This means that the model treats users with the same number of friends and same activity level as indistinguishable from each other when computing transitions between states, thereby ignoring additional individual differences. This simplification makes stochastic models tractable by reducing the number of parameters necessary to describe the population.
We determine transition rates between states by estimating the rate users {within each population} receive new posts from friends $\left(\Rpostsbar\right)$, the rate  they visit the site to view newly received posts $\left(\Rvisits(u)\right)$, the probability they view a post with given number of newer posts  $\left(\Pview(L)\right)$, and probability they find a viewed post sufficiently interesting to respond $\left(\Prespond(u)\right)$.
The next sections describe how we make these estimates from the available data.

Dependencies not included in the model's transition rates implicitly make independence assumptions. For example, we assume a user's response to one post does not affect the probability of response to subsequent posts, nor make the user more likely to look harder for future posts. Independence assumptions simplify the model and parameter estimation, though they may not be completely correct. Estimating transition probabilities depending on more parameters requires more detailed data than readily available for studies of social media.

\subsection{Parameters}
\subsubsection{Rate of receiving new posts}

A user $u$ receives posts at a rate determined by the number of friends and their posting frequency. For Twitter, this rate is approximately proportional to the number of friends~\cite{Hodas13icwsm}:
\begin{equation}\label{eq.received}
\Rreceived(u) = \Nfriends(u) \Rposts(\mbox{friends}(u))
\end{equation}
where $\Nfriends(u)$ is the number of friends user $u$ follows and $\Rposts(\mbox{friends}(u))$ is their average activity rate.

To accurately measure $\Rposts(\mbox{friends}(u))$
requires tracking a large number of users: not just the advocate and advocate's followers, but also the friends of those users, which is often impractical.
\remove{While such data is available, it is not useful for our study because we do not know when users visit Twitter to view their received posts (as opposed to when those users post new content). Thus a detailed time-history of the received posts for a user is not sufficient to determine the length of the user's list at the time of each visit, which is the relevant quantity for our stochastic model.}
Instead we consider a simplified model that averages over variations in received posts, so \eq{eq.received} becomes
\begin{equation}\label{eq.received avg}
\Rreceived(u) = \Nfriends(u) \Rpostsbar
\end{equation}
where $\Rpostsbar$ is the typical posting rate within a population of users.
Then, the main variation in the rate a user within a population receives new posts is due to variations in the \emph{number} of people that user follows, i.e., the number of friends, while taking the posting activity of each of those friends to be the typical population-level rate.
Since user activity rate varies considerably, we use the median for $\Rpostsbar$ as the typical value, rather than the mean, which is unreliable due to the long tail of the activity distribution.  \tbl{tbl.data} reports the value of $\Rpostsbar$ for each population of advocate followers we study. These values of typical activity rate are consistent with other studies of Twitter~\cite{Hodas13icwsm} which showed that, on average, users receive one post per day from a friend.


\subsubsection{Rate of visiting the site}

Data available from social media sites indicate when a user posts {or retweets some} content, but not when the user visits the site to look at existing content. Instead, we estimate the average rate at which users visit the site, $\Rvisits$, to be proportional to their posting rate:
\begin{equation}\label{eq.visits}
\Rvisits(u) = \ViewsPerPost \Rposts(u)
\end{equation}
where the constant of proportionality, $\ViewsPerPost$, is the average number of times a user visits Twitter to view content for each time the user posts. This ratio could vary among users: for the model we consider an average over the user population.

\remove{Any evidence for this proportionality from prior studies? I.e., evidence that the variation in visits is highly correlated with the variation in posts -- at least among users who post at least once -- those are the only users relevant for our data; users who just view Twitter without ever posting are not included.}

\subsubsection{Probability to view a post}

Social media produce far more content than users have time to examine. Thus there is a large chance that a user will never see a given post added to that user's list~\cite{Hodas12socialcom}. Precisely which received posts a user examines is not available in our data.
Instead, we combine \eq{eq.received avg} and \eq{eq.visits} with prior studies of how users navigate web sites to estimate whether a user views a post.

Consider a post $p$ from the advocate at time $t=0$. Let $L(t)$ be the number of posts the user who follows the advocate received (from all friends) during the time $t>0$ subsequent to receiving post $p$. If the user next visits the site at time $t$, the post $p$ will be at position $L(t)+1$ on the list of new posts.\footnote{We assume that user responds to the advocate's original post and not to any possible retweets of the post by one of user's friends. A retweet may boost the visibility of a post, but we ignore this effect for this study.}
\remove{This assumes Twitter uses a ``most recent first'' ordering -- do we need to comment on that as an approximation to using a ``most recent active'' ordering? Or is that not relevant to this discussion focusing on new content (in effect, any reordering due to activity becomes another contribution to the visibility parameter estimation of the unknown views to posts ratio.}

People are more likely to view items near the beginning of a list than those later in the list~\cite{huberman98,levene01,cooke08,Hodas12socialcom}. The ``law of surfing'' quantifies this behavior as arising from a biased random walk commonly applied to a variety of decision tasks~\cite{bogacz06}. This leads to an inverse Gaussian
distribution of the number of items $m$ a user views before stopping,
\begin{equation}\label{eq.stopping distribution}
e^{-\frac{\lambda  (m-\mu )^2}{2 m \mu ^2}} \sqrt{\frac{\lambda}{2 \pi m^3}}
\end{equation}
with mean $\mu$ and variance $\mu^3/\lambda$~\cite{huberman98}.
Thus the probability a user views a post, at position $L+1$ on the list, $\Pview(L)$, is the fraction of users who visit \emph{at least} $L$ items beyond the first one, i.e., the upper cumulative distribution of \eq{eq.stopping distribution}.

\remove{For the law of surfing parameters $\mu$ and $\lambda$ we use the same values for all users, in effect averaging over any systematic variation in how much of their lists of new content users examine during each visit.}

\begin{figure}
\centering   \includegraphics[width=\figwidth]{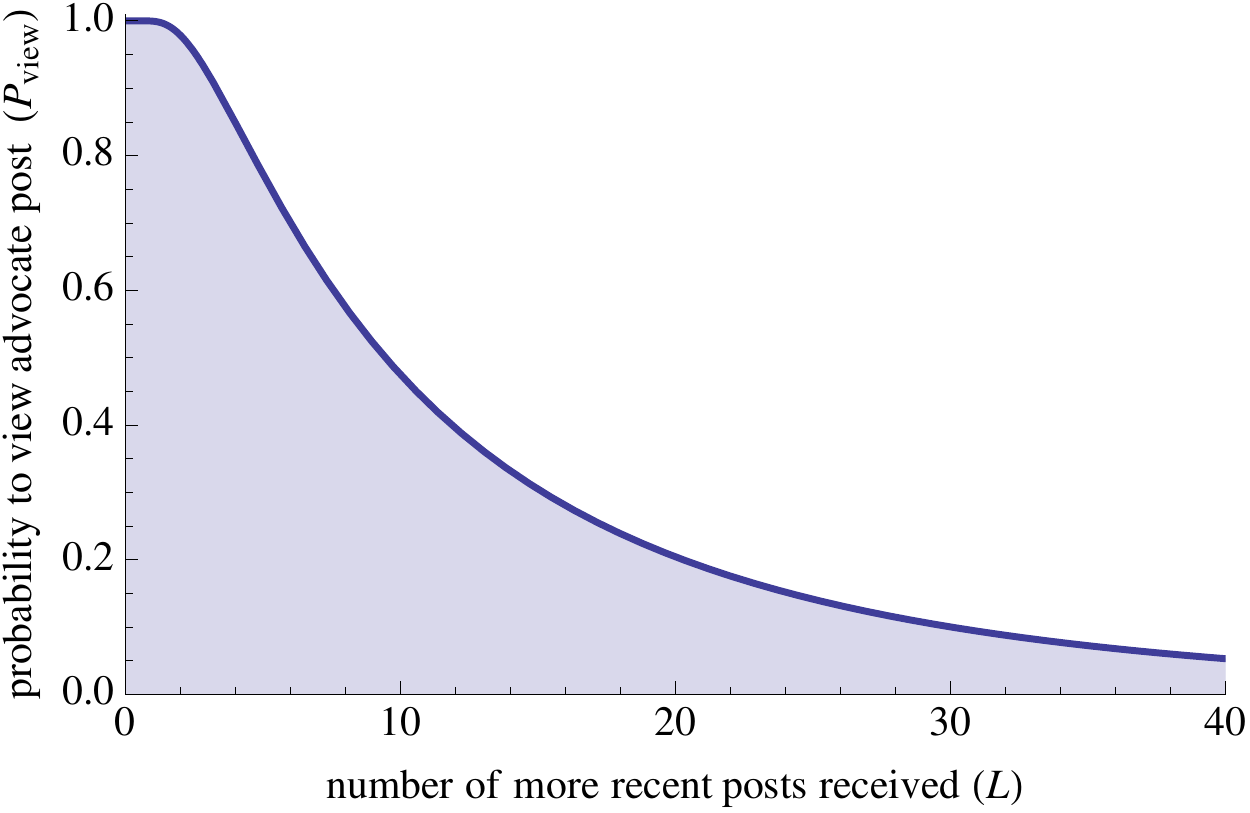}
\caption{Probability, $\Pview(L)$, a user views a post at position $L+1$ on the list using values $\mu$ and $\lambda$ given in \tbl{tbl.parameters}.
}\label{fig.law of surfing}
\end{figure}

Whether a user views a given post is a competition between how soon after that post the user visits Twitter and how rapidly the user's lists accumulates additional posts from friends. In general these rates vary over time as users become more or less active on the site. However, this competition depends only on the relative rates of these processes, leading us to approximate this competition as two Poisson processes with constant rates: $\Rvisits(u)$, the rate the user visits the site (\eq{eq.visits}), and $\Rreceived(u)$, the rate that user receives new posts (\eq{eq.received avg}).

This competition gives a geometric distribution for the number of additional posts $L$ the user received when first visiting after receiving a given advocate post:
\begin{equation}\label{eq.list length}
\Pposts(L|u) = \frac{1}{1+\rho}    \left(   \frac{\rho}{1+\rho}    \right)^L
\end{equation}
where $\rho = \Rreceived(u)/\Rvisits(u)$ is the ratio of rates the user receives posts to the rate that user visits Twitter. The expected number of new posts received by the time of the next visit is $\rho$.

Combining these factors, the probability the user views a given advocate post during the next visit to Twitter is
\begin{equation}\label{eq.visible}
\Pvisible(u) = \sum_L \Pposts(L|u) \Pview(L)
\end{equation}
The law of surfing, $\Pview(L)$, decreases rapidly with position $L$, as shown in \fig{fig.law of surfing}. Thus $u$ typically will only see the advocate's post if they visit Twitter relatively soon after the advocate posts. That is, most of the contribution to $\Pvisible(u)$ is from the first few terms of the sum.

\remove{Possibly discuss whether user might see the post at a later visit after not seeing it during the first visit to Twitter after the advocate's post. I.e., the model gives the user just one chance to see the post -- during the next visit to Twitter. Since subsequent visits will have the post even further down the list and law of surfing decreases rapidly with position on the list, this is a reasonable approximation (and, presumably, just amounts to a small rescaling of the parameters in the model).}

\remove{
Assuming a Poisson process for user visits, the average time after a post from the advocate until the user's next visit is $1/\Rvisits(u)$. During this time, that user receives additional posts at a rate $\Rreceived$ given by \eq{eq.received avg}. Thus when the user visits, there are
\begin{equation}
L \approx \Rreceived / \Rvisits
\end{equation}
newer posts in the user's list and the probability the user sees the advocate's post is $\Pvisible(u)=\Pview(L)$.
} 

\begin{figure}
\centering   \includegraphics[width=\figwidth]{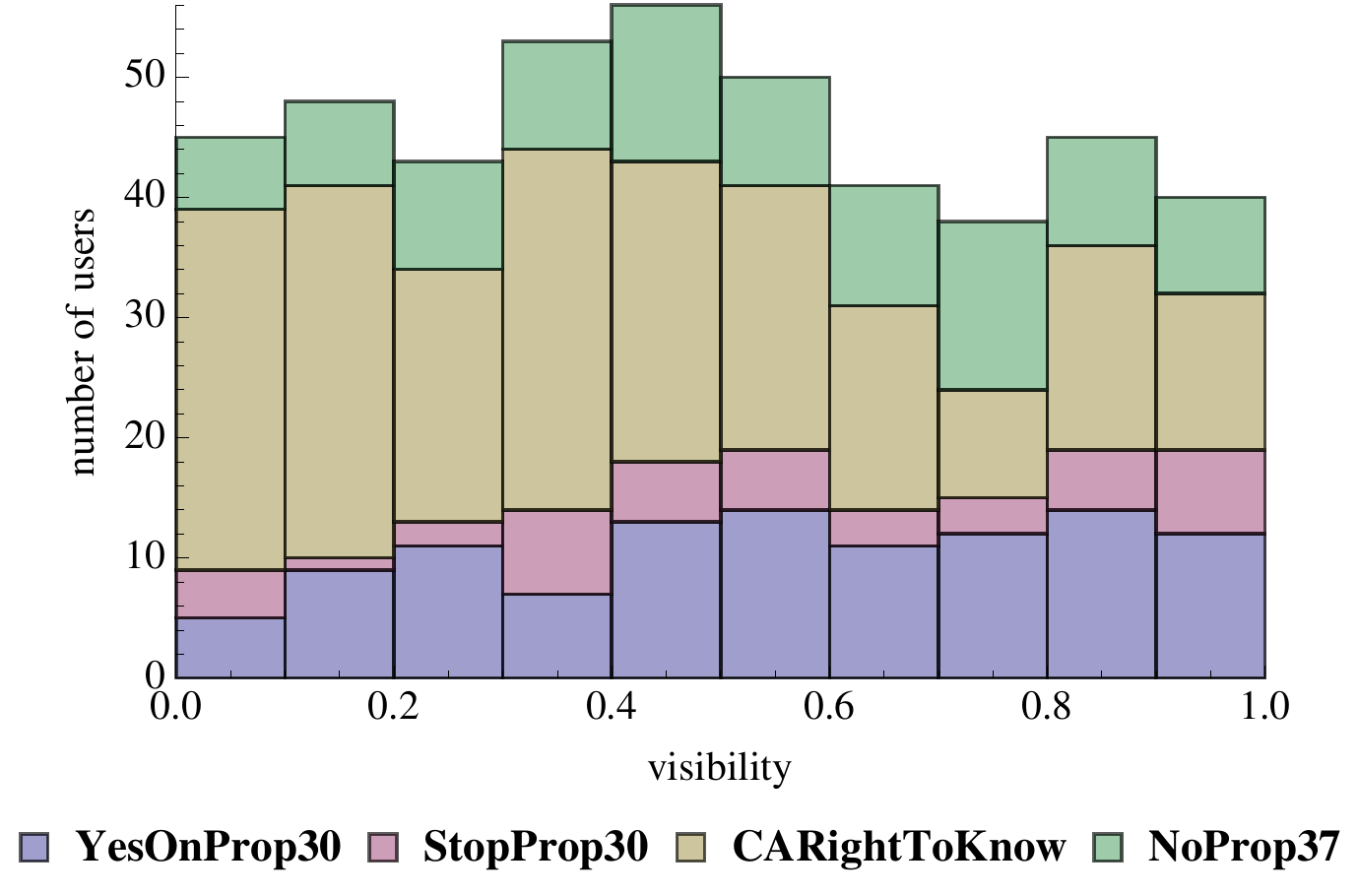}
\caption{Distribution of $\Pvisible$ among users from different populations, using model parameters given in \tbl{tbl.parameters}.
}\label{fig.visibility}
\end{figure}

Evaluating visibility $\Pvisible(u)$ (\eq{eq.visible}) involves the law of surfing parameters $\mu$ and $\lambda$, and the ratio of views to posts, $\ViewsPerPost$. \fig{fig.visibility} shows the distribution of $\Pvisible$, showing the values span the range from 0 to 1 (and are consistent with a uniform distribution, Kolmogorov-Smirnov test $p$-value is $0.10$). This wide variation in visibility means models of how users respond to posts should account for visibility, not just the user's activity or interest in the topic.

\subsubsection{User response to a post}

After reading a  post, the user may respond. Social media provide various ways for users to respond. These include rating the post (e.g., voting on Digg), forwarding it to followers (e.g., by retweeting it) or commenting on it (e.g., writing a book review on Amazon). Users can also take action outside of the social media setting, e.g., forwarding by email or donating time or money to a cause based on the post. The likelihood of a response depends on the nature of the response considered: it requires more effort to write a book review than to forward a post to another user.

With this variety of ways for users to respond, the particular definition of response we focus on is somewhat arbitrary. However, for testing the model we need a response that is available in our data. For the purpose of this study, the content we consider is posts from a campaign advocate and responses are followers who retweet those posts.

This choice of response has two benefits beyond availability of data. First, posts from a campaign advocate are likely more focused on the single topic of the campaign, leading to a more homogeneous set of posts than general content. This should improve the applicability of our model's assumption of averaging behavior over the content we consider.
Second, in the context of campaign advocacy, spreading the advocate's message to more people, with at least an implicit endorsement of the user, is likely to be an explicit goal of the advocate. This motivates our defining a response as a follower reposting a tweet from the advocate. Since our data includes all tweets of the followers during the period under study, we know if a follower responded in this way.
\remove{determined as described in Data section -- presumably perfect accuracy by using structure of the tweet; not the estimate of other properties based on tweet content analysis}

\subsubsection{Probability a user responds to a viewed advocate post}

The model characterizes user response by the probability the user finds the post sufficiently interesting to warrant a response. We estimate the level of interest each follower $u$ has in the posts, $\Pinteresting(u)$, as a product of two quantities: the degree of a user's interest in the topic as a whole, $\Ptopic(u)$, and the conditional probability of interested users to respond to a post from the advocate, $\Pact(u)$, in this case by forwarding the post to the user's followers.

We estimate a user's overall interest in the campaign topic, $\Ptopic(u)$, from the number of posts from that user and the number of those posts related to the topic, as determined by content analysis of $u$'s posts (described Section~\ref{sec.data}).
\remove{check that the Data section does describe this; possible reference by section number instead of name, by adding a section label to Data}
A simple estimate takes $\Ptopic(u)$ to be the fraction of the user's posts on the topic. However, this doesn't account for the confidence in this estimate. For example, a user with 2 out of 3 posts on the topic and a user with 20 out of 30 posts on the topic both have the same fraction; but the user with more posts gives more confidence in the estimate.

To account for this variation, we consider a {\it distribution} of possible $\Ptopic(u)$ values for each user. Specifically, if a user has $n$ posts, the probability $m$ of them are on the topic, assuming independence, is the binomial distribution $\Binomial(n,\Ptopic;m)$ where
\begin{equation}\label{eq.binomial}
\Binomial(n,p;m) = \binom{n}{m} p^m (1-p)^{n-m}
\end{equation}
Our data provides the values for $m$ and $n$ for each user, so instead of treating this as a distribution for $m$ given $n$ and $\Ptopic$, we have a beta distribution for the possible values of $\Ptopic$ based on the observed values of $m$ and $n$:
\begin{equation}\label{eq.beta}
\Prob(\Ptopic) = (n+1)\, \Binomial(n,\Ptopic;m)
\end{equation}
where the factor $n+1$ ensures the distribution is normalized: $\int_0^1 \Prob(\Ptopic) \,d\Ptopic=1$.

When the user has many posts (i.e., large $n$), the distribution is narrow and $\Ptopic \approx m/n$, the fraction of the user's posts that are on the topic. But for a user with few posts, the distribution is broad, reflecting the uncertainty in estimating the user's interest from just a few posts.

The second factor determining a response is the probability an interested follower $u$ responds to a particular advocate post  after seeing it, $\Pact(u)$. In our data, users who we determine are opposed to the advocate's position generally do not respond to the posts, so we take $\Pact(u)=0$ for such users.
\remove{Caveat: the stance is an estimate -- if a supporter is incorrectly classified as an opponent and happens to respond, using $\Pact(u)=0$ will give $-\infty$ likelihood, so could be a problem for max. likelihood based estimation. More generally, we could estimate $\Pact$ for user we estimate to be non-supporters. We will find $\Pact=0$ as the max. likelihood result if non-supporters in the training data never respond, so get the same result but described in a more principled way}
For the rest, ``interested supporters,''  we use the average over these users $\Pact = \langle \Pact(u) \rangle$.
We thus consider users to have differing levels of interest in the topic while assuming content from an advocate is fairly homogeneous in leading interested users to respond.

Combining these factors gives
\begin{equation}\label{eq.interest}
\Pinteresting(u) = \Ptopic(u) \Pact
\end{equation}
as our estimate of a user having sufficient interest in an advocate's post to respond to it, conditioned on that user's value of $\Ptopic$.

\remove{A possible improvement: use community response to a specific post to estimate its average interest to ``interested supporters'' (assuming average visibility). That is, make $\Pinteresting(u,p)$ a function of both the user $u$ and post $p$ [in contrast to our Digg study where interestingness depended on the story $p$ but not the user (other than in broad groups, such as submitter fans)]. But then we need to decide how to combine that average per-post interest with user-specific differences, estimated by overall interest in topic, to get an estimate of user-specific interest in each advocate post. And for testing, we need some way to estimate interestingness of a particular advocate post from a training set, which would require picking training cases from all the advocate's followers instead of split training/testing from different advocates.}

\subsubsection{Response distribution}

The model gives the probability a user $u$ responds to an advocate's post as the product of the probability the user sees the post and the probability of finding it sufficiently interesting once viewed:
\begin{equation}\label{eq.respond conditional}
\Prespond(u|\Ptopic) = \Pvisible(u) \Pinteresting(u)
\end{equation}
conditioned on the value of $\Ptopic$ for the user. For a situation with $N$ advocate posts, assuming a user makes independent choices to view and respond to each post, the probability that user responds to $M$ of the $N$ advocate posts is binomially distributed: $\Binomial(N,\Prespond(u|\Ptopic);M)$.

Integrating over the distribution of $\Ptopic$ for that user (\eq{eq.beta}) gives the distribution of the number of responses by that user, $\Prespond(u)$:
\begin{eqnarray}\label{eq.respond}
\int_0^1 \Prob(\Ptopic)\; \Binomial(N,\Prespond(u|\Ptopic);M) \; d \Ptopic \nonumber & &\\
= \binom{M+m}{m} \binom{N}{M} \binom{M+n+1}{M}^{-1} \mathcal{L}(u) & &
\end{eqnarray}
where $\Prob(\Ptopic)$ is the distribution of $\Ptopic$ for the user (\eq{eq.beta}) and
\begin{equation}\label{eq.likelihood}
\mathcal{L}(u) =  \mathcal{A}^M \; {_2F_1}(m+M+1,M-N;M+n+2;\mathcal{A})
\end{equation}
where $\mathcal{A} = \Pvisible(u) \Pact$\remove{, so that $\Prespond(u|\Ptopic) = \mathcal{A} \Ptopic(u)$ from \eq{eq.interest} and \eq{eq.respond conditional},} and ${_2F_1}$ is the hypergeometric function~\cite{abramowitz65}.


Evaluating the response distribution for a given user requires the user-specific data given in \tbl{tbl.user data}. The table includes two parameters that are the same for all users in a population, i.e., followers of an advocate. These are the number of advocate posts, $N$, and the typical activity rate for users in that population, $\Rpostsbar$. \tbl{tbl.data} gives the values for these population-specific parameters.

\begin{table}
\centering
\begin{tabular}{ll} \hline
\multicolumn{2}{c}{{\it user-specific parameters}} \\
posting activity rate	  	& $\Rposts(u)$ \\  
number of friends		& $\Nfriends(u)$\\
stance on topic (boolean)			& same as the advocate?\\
number of posts on topic	& $m$\\
number of posts		& $n$\\
number of advocate posts retweeted 	& $M$\\ \hline
\multicolumn{2}{c}{{\it population-specific parameters}}\\
number of advocate posts				& $N$\\
typical posting rate					& $\Rpostsbar$\\
\end{tabular}
\caption{User data.}\label{tbl.user data}
\end{table}

\begin{figure}
\centering   \includegraphics[width=\figwidth]{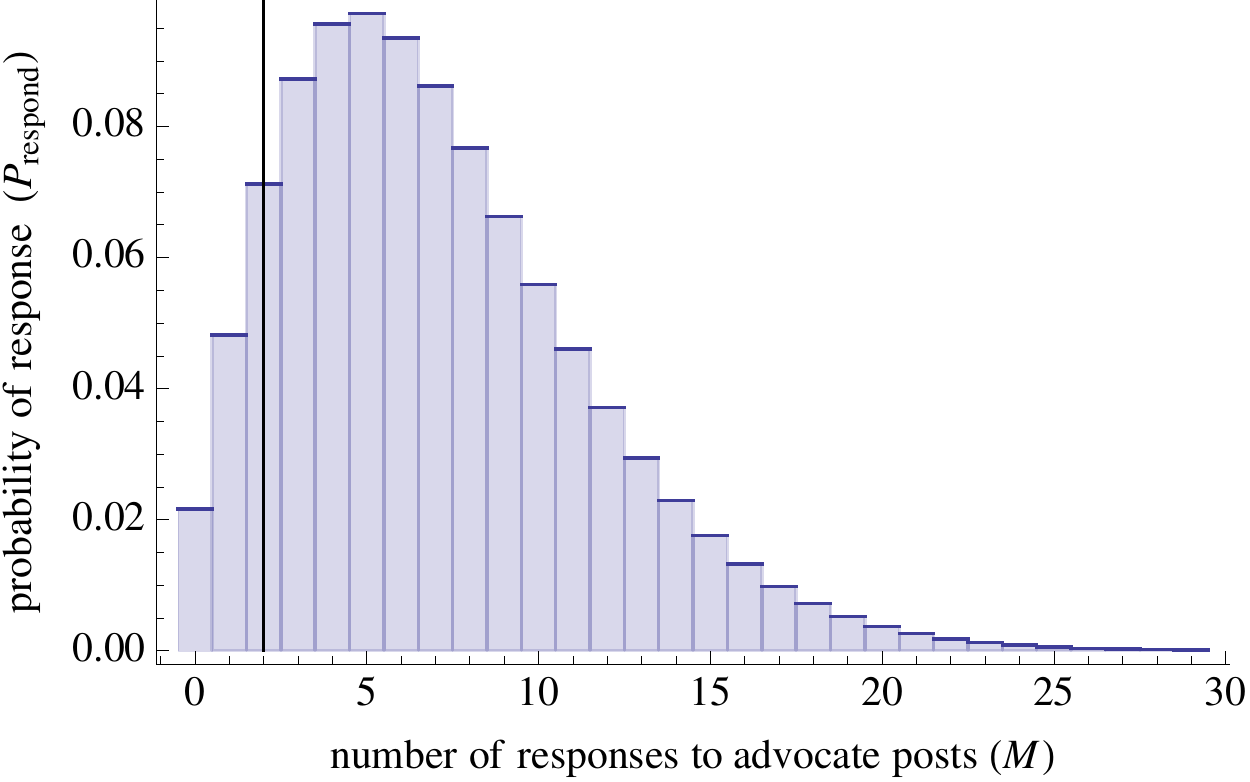}
\caption{Distribution of response $M$, i.e., $\Prespond$ as a function of $M$ from \eq{eq.respond}, for a user, using model parameters given in \tbl{tbl.parameters}. The vertical line shows this user's actual response, $M=2$.
}\label{fig.response}
\end{figure}

\fig{fig.response} is an example of the probability distribution for responses, $\Prespond$, for a user. In this case, the user's actual response ($M=2$) is among the more likely number of responses predicted by the model. This figure shows that the model produces a distribution over the possible outcomes, providing both an estimate of the likely user response and the confidence in that estimate from the width of the distribution.

\subsection{Parameter estimation}

We estimate values of the model parameters by maximum-likelihood: the choices of model parameters that maximize the probability of the observed responses from a set of users according to the model. We treat the users as responding independently so the probability of the observed responses is the product of $\Prespond(u)$ for each user, from \eq{eq.respond}.

For this maximization, we only need factors that depend on the model parameters, i.e., the value of $\mathcal{A}$ appearing in \eq{eq.likelihood}. All other values are known from the user data. For convenience, we take the logarithm of this expression, thereby maximizing
\begin{equation}
\sum_u \log( \mathcal{L}(u) )
\end{equation}
where the sum is over all users $u$ in the training set, which we take to be the followers of @YesOnProp30. The followers of the remaining advocates listed in \tbl{tbl.data} are our test set to evaluate the model.

Our model does not determine the law of surfing parameters beyond the requirement that $\mu \approx \lambda$ to maximize the likelihood. This is because the likelihood is nearly the same if users visit Twitter twice as often but only look at half the number of items on their list, i.e., rescaling $\ViewsPerPost \rightarrow 2 \ViewsPerPost$ while dividing $\mu$ and $\lambda$ by 2.
Thus for this study we use the law of surfing parameters determined for a prior study of social media~\cite{hogg12b}, multiplying by 15 to convert from ``page number'' to ``item number'' on the list of new posts. These parameters have $\mu \approx \lambda$, so are consistent with our model for Twitter.
\remove{Although we don't have a strong reason to believe people view Digg stories the same as posts from friends, without clear data to the contrary, reusing Digg parameter estimates improves claims of generality of the stochastic approach. On the other hand, estimating law of surfing parameters from training set of campaign data and seeing those estimates are close to prior study of Digg gives confidence in the estimation for the campaign data -- unfortunately the model only gives a loose constraint that $\mu \approx \lambda$, which holds for the Digg parameters.}

\tbl{tbl.parameters} shows the maximum-likelihood estimates of the model parameters, with ranges indicating 95\% confidence intervals.

\begin{table}
\centering
\begin{tabular}{ll}
\hline parameter & value \\ \hline
law of surfing	  	& $\mu=14 \pm 0.6$ \\  
				& $\lambda=14 \pm 1.8$ \\
views per post		& $\ViewsPerPost=38 \pm 20$\\
response by interested supporter	& $\Pact = 0.12 \pm 0.03$\\
\end{tabular}
\caption{Model parameters estimated from data about population of followers of @YesOnProp30.}\label{tbl.parameters}
\end{table}

\section{Results}
We evaluate the model on a test set consisting of the followers of the second through fourth advocates in \tbl{tbl.data}. Thus we use model parameters estimated for one advocate (@YesOnProp30) to predict follower behavior for three other advocates. We compare predictions of the stochastic model with a baseline regression model, which is described next.


\remove{This choice of training/test set stresses our point that we are learning on one population and applying the model to others. There is interest in machine learning community in "transfer learning", and this is an example of that.  (Is there a relevant citation for transfer learning?)}

\subsection{Regression model}
\label{sec.regression}

For comparison with the stochastic model, we consider a logistic regression model relating overall user activity to response to advocate posts. Unlike the stochastic model, a regression model does not consider the user as transitioning through a series of states to decide on a response. Instead, it simply considers users who are more active on the site are also more likely to respond.

Since user activity varies over a wide range and has a long tail, we find a regression on the log of the activity rate, i.e., $\log \Rposts(u)$, provides better discrimination of response than a using $\Rposts$ itself.
Determining the fit based on the same training set as used with the stochastic model, i.e., the followers of @YesOnProp30, we find
\begin{equation}
\Prespond(u) = \frac{1}{1+\exp(-(\beta_0 + \beta_1 \log(\Rposts(u))))}
\end{equation}
with $\beta_0 = -5.01 \pm 0.06$ and $\beta_1 = 0.11 \pm 0.03$.

\subsection{Predicting user response}
For testing, the actual response $M$ is the dependent variable predicted by the model. Specifically, the model gives the distribution of $M$ for a user (\eq{eq.respond}) based on the model parameters (\tbl{tbl.parameters}), and other data for that user (i.e., the values in \tbl{tbl.user data} other than $M$).
We use the expected value from this distribution as the model's prediction of $M$.

\begin{table}[!htb]
\centering
\remove{ 
\begin{tabular}{lc}
\hline
{\it model}		& {\it correlation}  \\ \hline
stochastic 	&  0.46 	 \\
regression	&  0.24 	 \\
\end{tabular}
}
\begin{tabular}{lccc}
\hline
				& \multicolumn{2}{c}{{\it model}} & \\
{\it advocate}		& {\it stochastic}  & {\it regression} & {\it same?} \\ \hline
@YesOnProp30 		&  0.44 	 	& $0.09^*$	& 0.005 \\
@StopProp30		&  $0.20^*$	& 0.35		& 0.46	 \\
@CARightToKnow 	&  0.29		&  $-0.08^*$	& $10^{-4}$	 \\
@NoProp37			&  0.60 		& $-0.11^*$	&  $10^{-7}$\\
\end{tabular}
\caption{Spearman rank correlation between model prediction and observed follower response. Last column is $p$-value of Spearman rank test for whether the two models have the same correlation. Asterisks indicate correlations not significantly different from zero at $5\%$ $p$-value by Spearman rank test.}
\label{tbl.correlation}
\end{table}

\tbl{tbl.correlation} shows the Spearman rank correlation between the predicted and observed number of responses for users in each test set.
Note that the first line reports results of testing on the data used for training the model. The last column reports the $p$-value of a statistical test to identify differences between the models. The closer this value is to zero, the more confidence this gives for rejecting the hypothesis that the two models actually have the same correlation and just produced the observed difference in correlation on these samples of users by chance.
Overall, the stochastic model has a larger correlation except for @StopProp30, where the relatively small number of followers and advocate posts (see \tbl{tbl.data}) are not sufficient to identify differences from the regression model.

\remove{The prediction discussion focus is on the expected value of $M$ compared to the observed value. We could discuss other measures that include the variance.
Examples:
1) Compare prediction error to standard deviation: The two models are not that much different in the overall size of errors, with the stochastic model tending to have more cases of extreme errors while doing better than the regression model on cases with small errors.
2) How often the observed value is, say, within 95\% portion of the distribution (we'd expect about 5\% of the users to be outside this range, but there are actually more extreme cases than expected, indicating additional variation among users not included in the model).
}


The model provides a distribution of responses, not just a single prediction. This additional information from the model is an indication of the accuracy of the prediction. \tbl{tbl.error correlation} shows one characterization of this accuracy: the Spearman rank correlation between prediction error and the standard deviation of the distribution returned by the model. The prediction error is the absolute value of the difference between the expected value of the distribution and the observed number of responses to the advocate posts. The stochastic model has a larger correlation except for @StopProp30. Correlations are larger than those of the prediction itself (\tbl{tbl.correlation}). We find that while variability of user behavior not included in the model gives many relatively large prediction errors, the standard deviation of the model nevertheless provides a good ranking of the prediction accuracy.

\begin{table}[!htb]
\centering
\begin{tabular}{lccc}
\hline
				& \multicolumn{2}{c}{{\it model}} & \\
{\it advocate}		& {\it stochastic}  & {\it regression} & {\it same?} \\ \hline
@YesOnProp30 		&  0.76 	 	& 0.33		& $10^{-5}$ \\
@StopProp30		&  0.85		& 0.81		& 0.54	 \\
@CARightToKnow 	&  0.95		&  0.49		& 0 \\ 
@NoProp37			&  0.93 		& 0.57		&  0 \\ 
\end{tabular}
\caption{Spearman rank correlation between prediction error and standard deviation of model distribution.
}
\label{tbl.error correlation}
\end{table}

The Spearman rank correlation values for training and testing the model on the same population of users (@YesOnProp30) is similar to the correlation values when training on one population (@YesOnProp30) and testing on the other populations (@StopProp30, @CARightToKnow,@NoProp37). This highlights the power of stochastic models to learn from one campaign and transfer to another, despite differences between individual campaigns, including those due to the issues involved, campaigner style, follower preferences, etc.

\subsection{Classification of user response}

In addition to using the model to predict how many advocate posts a user will respond to, models can classify users by their relative response among all the advocate's followers. For instance, models can identify the subset of users who are likely to respond the most to advocate posts rather than precisely predicting how often they will respond. This classification of users is analogous to classifying content on social media sites, e.g., distinguishing stories likely to get many or few votes rather than predicting the precise number of votes~\cite{hogg12b}. Such predictions form the basis of using crowd sourcing to select a subset of submitted content to highlight~\cite{Lerman08wosn}.

As an example, we apply the stochastic model to predict which users in the test set will be among the top 25\% of responders, measured by the fraction of advocate posts they responded to. One way to use a model for this classification task is to select the users whose predicted response is among the top 25\% of those predictions. Model performance on this classification task is then the extent these selected users correspond to the actual top responders. \tbl{tbl.classify} shows the fraction of users in the test set incorrectly classified by this procedure, i.e., the fraction of users predicted to be among the top 25\% who were not, or vice versa. The table also shows the precision and recall of this classification, i.e., fractions of predicted top responders who are, and fraction of top responders who are predicted to be so, respectively. In this case, classification by the stochastic model is significantly better than a random classifier, i.e., randomly selecting 25\% of the users. On the other hand, the regression model is consistent with random classification.

\begin{table}[!htb]
\centering
\begin{tabular}{lcccc}
{\it model}		& {\it error fraction} & {\it precision} & {\it recall}  & {\it random?} \\ \hline
stochastic 	&  30\% 	& 40\%	& 40\%	& $10^{-4}$ \\
regression	&  36\% 	& 27\%	& 27\% 	&  0.2\\
\end{tabular}
\caption{Classification of top 25\% responders. The last column is the $p$-value of this error rate arising from a random classifier according to the Fisher exact test of proportions~\cite{collett03}}\label{tbl.classify}
\remove{Should we desegregate these values? Since \fig{fig.precision recall} shows aggregate behavior (and we don't have room for separate figures for each advocate), perhaps better not to take additional space to show individual advocates, especially since this table has 8 numbers for each case we decide to show. We are adjusting somewhat for each advocate by classifying based on the \emph{fraction} of responses rather than actual number: this normalizes for the different numbers of advocate posts among the 3 advocates we use for testing.}
\end{table}

\begin{figure}
\centering \includegraphics[width=\figwidth]{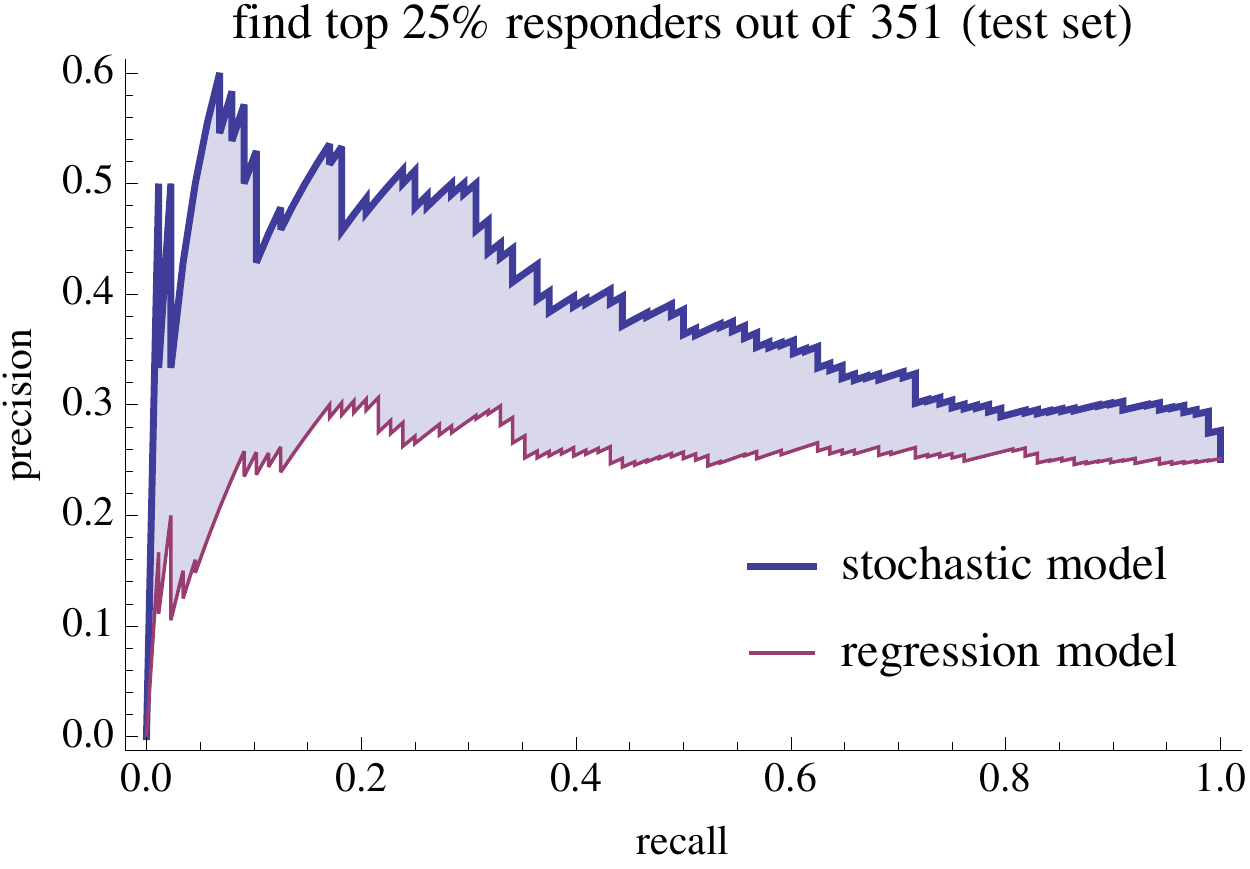}
\caption{Precision vs.~recall for identifying top responders.}\label{fig.precision recall}
\end{figure}

A more general measure of classification performance is the precision vs.~recall curve, shown in \fig{fig.precision recall}. For each model, we sort the $U=351$ users in the test set according to their predicted fraction of response to the advocate's posts. For $k=1,\ldots,U$, we examine the set of users with the $k$ largest predicted responses. For example, when $k=1$ the set is the single user with the largest predicted response, and when $k=U$ the set has all users. As $k$ increases, the figure shows the fraction of these $k$ users that are among the observed top 25\% responders (precision) vs.~the fraction of the observed top responders included among the $k$ users (recall). Better classifiers have higher curves in this figure: able to identify a large fraction of the actual top responders without also including many less responsive users. By comparison, random selection of users would give, on average, $25\%$ precision for any value of recall. The curve for the regression model does not differ significantly from this precision value.

A specific classifier using this procedure amounts to selecting a position on the precision-recall curve by picking the number of users to consider as the top responders, i.e., a value of $k$. For instance, the values in \tbl{tbl.classify} correspond to $k=88$, i.e., 25\% of the 351 users in the test set. In practice, selecting a position on the curve, i.e., choice of $k$, depends on how important it is to avoid false positives vs.~false negatives. In the example discussed in \fig{fig.precision recall}, with somewhat smaller values of $k$, the stochastic model gives higher precision.

\subsection{Estimating user interest}

Both stochastic models and simpler regression models can classify users based on their likely response. However,  the stochastic model can also estimate underlying state transition probabilities for users. For instance, this could distinguish users who do not respond mainly due to visibility (e.g., users who follow many others or do not visit Twitter often) from those who do not respond due to lack of interest. The former group, i.e., interested users who do not see the advocate posts, would be more likely to respond to higher-visibility messages (e.g., direct mail) than the latter group. Thus this classification based on \emph{why} users are not responding could help the advocate focus limited resources on reaching supportive users. Since the model characterizes user interest by a distribution of values, the model can not only suggest users with high interest but also estimate the confidence in that assessment based on the width of the distribution.

\begin{figure*}
\centering
\begin{tabular}{cc}
\includegraphics[width=\figwidth]{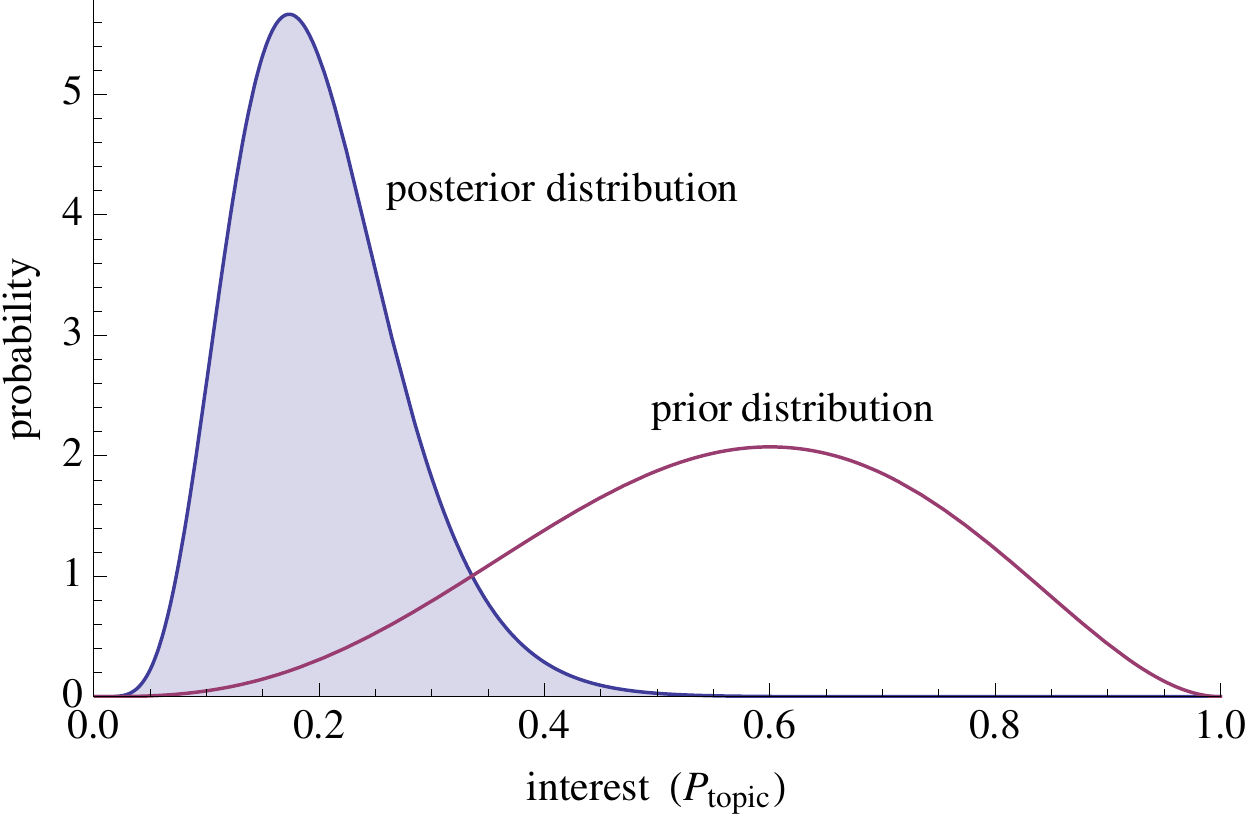} &
\includegraphics[width=\figwidth]{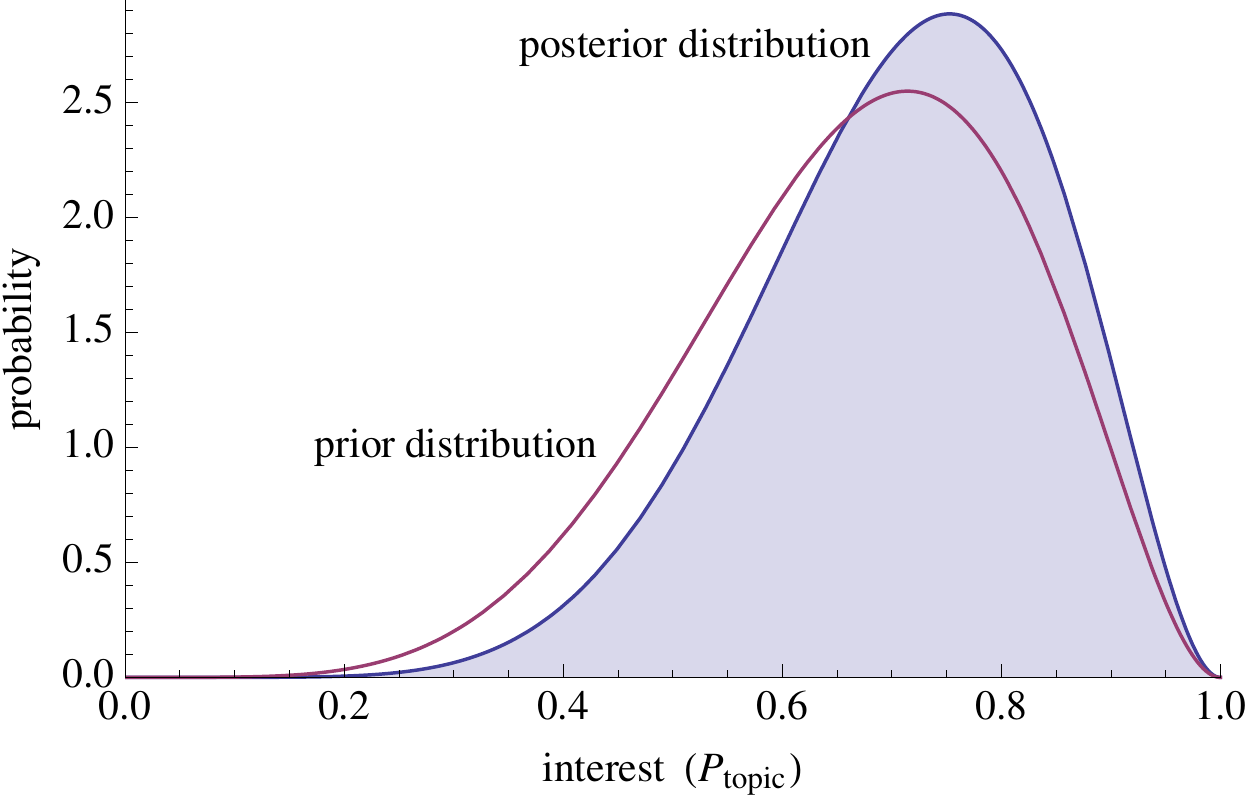} \\
(a) low interest user & (b) high interest user
\end{tabular}
\caption{Estimated prior distribution of $\Ptopic$ values for two users and posterior distribution after observing how the users respond to advocate posts. (a) Low interest user with 3 out of 5 posts on topic, who responds to 3 of 391 advocate posts.
(b) High interest user with 5 out of 7 posts on topic, who responds to 2 of 391 advocate posts.}\label{fig.interest distribution}
\end{figure*}

As an example, \fig{fig.interest distribution} shows how the {visibility-adjusted} prediction alters the estimated likely range of user interest after observing how that user responds. The prior distribution (\eq{eq.beta}), before observing responses to advocate posts, is from a content analysis of the user's posts. For the user in \fig{fig.interest distribution}(a), the values are $m=3$ and $n=5$, which suggests a relatively high interest in the topic, but with large uncertainty due to the small number of posts by that user. After observing the user responses to advocate posts (in this case, responding to $M=3$ of $N=391$ posts), the posterior distribution for $\Ptopic$ is proportional to the integrand of \eq{eq.respond}. This combines the prior distribution with response due to a combination of visibility and interest to give the posterior distribution of $\Ptopic$ shown in the figure. In this case, the posterior distribution suggests a user with relatively low interest.

By contrast, \fig{fig.interest distribution}(b) shows the estimates for an apparently similar user: $m=5$ and $n=7$ and responding to only $M=2$ out of $N=391$ advocate posts. In this case, the model accounts for the low response as due to low visibility of the advocate's posts and estimates the user's interest in the topic is likely a bit higher than expected from the prior content analysis of the user's posts. Thus, according to the model, this user is likely far more interested in the topic than suggested by the low response to advocate posts.

To test the predicted difference in interest between these users, we examined additional data on their posts. The first user is a news reporter from outside California who follows only one advocate, namely @NoProp37, and focuses on national politics rather than California state politics. The second user follows 25 advocates, including 10 who post on proposition 37, on both sides of the issue. Thus the second user appears significantly more interested in the topic than the first, consistent with the model prediction shown in \fig{fig.interest distribution}.

\subsection{Discussion}

We find that both prediction of response and classification are better when we account for transitions among user states involved in social media than using a statistical regression based on overall activity.
In particular, even with crude estimation of the model properties, using the state-based model leads to better predictions. In practice, data from social media often do not include all relevant details of user behavior (e.g., when they visit the site or view particular content). Thus, this study shows stochastic models can give reasonable performance in practice despite using coarse estimates based only on readily observed behavioral data.

However, the relatively low absolute performance for the stochastic model shows that user-specific details not accounted for are important for improving accuracy.
For instance, the rate of receiving new posts depends on the activity rate of a user's friends, $\Rposts(\mbox{friends}(u))$, which we approximated by $\Rpostsbar$, the typical posting rate within a population of users.  The distribution of user activity varies considerably, having a long tail.  Some users are quite prolific, while others rarely post.  Thus data on the activity of each user's friends could improve the model by providing $\Rposts$ on a user-specific basis.

Another model parameter,  $\Pact(u)$, estimates a user's likelihood to retweet interesting advocate posts.  We took this value to be the average over all users, $\Pact$.  However,  people use retweets differently.  Some users mainly retweet posts rather than posting original material, indicating a higher probability of responding to a viewed tweet.  Other users have almost no retweets, suggesting these users prefer not to rebroadcast posts. This user-specific variation could contribute to the difference between the prior and posterior distribution seen in \fig{fig.interest distribution}.   A possible extension of our model is to use each user's propensity to retweet in general as part of estimating $\Pact(u)$ for that user.

Not every post by an advocate is of interest to all of the followers, such as a post directed to another user. Therefore, a user's interest in a tweet depends on the post's appeal  and the individual's interest in the topic.  A direction for future work is to use the overall response to the tweet as an indication of its appeal.  The model would then combine both the user's estimated topic interest and the community's overall interest in each tweet to determine the user's interest in that particular tweet.
By addressing the tweet-specific interest, this could improve $\Pact$ by including a measure of the \emph{content}, i.e., the appeal of a post, instead of assuming all advocate posts are equally appealing, on average.  This contrasts with the improvement to $\Pact$ mentioned above, which focuses on differences in a \emph{user's} tendency to respond in general.

\section{Related Work}
Stochastic models are the basis of population dynamics models used in a variety of fields, including statistical physics, demographics, epidemiology, and macroeconomics. In the context of social media, stochastic models identify mechanisms relating the design of social media sites to the behavioral outcomes of their users.
Previous applications of the stochastic modeling framework focused on describing the aggregate behavior of many people by {\em average} quantities~\cite{Iribarren09,social-physics,Lerman12tist}, such as the average rates at which people  contribute content or respond to emails, and so on. 
A series of papers applied the stochastic modeling framework to social media, specifically to examine the evolution of popularity of individual items shared on the social news aggregator Digg~\cite{Lerman07ic,hogg09c,hogg12b,Lerman12tist}. These studies assumed the user population had similar interest in each item, or similar when separated in broad populations (e.g., follower or not of the item's submitter), whereas the quality of each shared news item varied, leading to differences in their eventual popularity.  In this paper, by contrast, we apply the models to describe and predict the behavior of an individual user, averaging over any differences in quality of content (i.e., the advocate's posts). 

When inferring response of users to items shared on social media, computer scientists generally consider only user's interest in the topics of an item~\cite{Lauw12,WangB11,purushotham}, or the number of friends who have previously shared the item~\cite{Myers12kdd}. However, as we have demonstrated in this paper, response is conditioned on both interest and visibility of the item, so that a lack of response should not indicate a lack of interest.
Failing to account for the visibility of items can lead to erroneous estimates of interest and influence~\cite{Krumme12}. Stochastic modeling framework allows us to factor in the visibility of items in a principled way.

Stochastic models are similar to Hidden Markov Models (HMMs) frequently used to model systems in which the observed behavior of an individual is a result of some unobserved (or hidden) states (see, e.g., \cite{Raghavan2012AOAS}). In contrast to HMMs, the goal of which is to identify hidden states that maximize the probability of an observed action sequence, stochastic models are best suited for predicting the evolution of the observed actions of a population of individuals. Population-level analysis, combined with relevant independence assumptions, enables stochastic models to parsimoniously describe observed behavior of many individuals, and also leads to more tractable parameter inference.

\section{Conclusion}

We showed that a stochastic model of user behavior gives useful predictions even when available data does not include aspects of user behavior most relevant for estimating model parameters. Instead, we used readily available proxy estimates.
We attribute success of the stochastic modeling approach to capturing relevant details of user behavior in social media that, although unobserved, condition the observed behavior. Perhaps the most important aspect of behavior our stochastic model accounts for is visibility of the post, which depends on how many newer posts are above it on the user's list, and how likely the user is to scan through at least that many posts. We demonstrated that accounting for visibility allows us to more accurately predict response and determine a user's actual interest in the topic. We see performance gains even with simple parameter estimates based on population-wide averages, suggesting the possibility that a model that better accounts for user heterogeneity will produce even better quantitative insights.

\remove{Discuss success of model in spite of what may appear to be significant limitations on its use due to data limits. Long tails, extreme heterogeneity of user population and content quality. Suggests need for detailed data on user behaviors to apply model. Or that simplification based on averaging would be so poor an approximation as to prevent use of the stochastic model. Even with crude estimates of user behavior and averaging over the long-tail distributions (a situation that suggests mean-field approximations for stochastic models are not useful), we find useful predictions from the stochastic model. Thus even with simple, approximate and limited data on user behavior, the model provides useful quantitative insights.}

The applications highlighted in this paper do not exhaust the list of potential applications of stochastic models.
Additional applications include predicting response to new posts from early user reactions (e.g., as with Digg~\cite{hogg12b}),
\remove{identifying highly responsive users based on responses to multiple posts after accounting for visibility (these could be good users to help with campaign), }
and using visibility-adjusted response to determine most useful or compelling campaign messages to show to new users (i.e., a ``visibility-adjusted'' popularity measure).

Our treatment considers user parameters as static. However, user's activity, interest in the topic and willingness to support the advocate could change in time.  The stochastic model could incorporate changes by re-estimating parameters over time. This could provide useful feedback to the advocate to gauge reaction to specific campaigns. In particular, by separating out the effects of visibility, the model could indicate how the level of interest changes after the campaign.

\subsection*{Acknowledgements}
\noindent We thank Suradej Intagorn for collecting the data used in this study.
This material is based upon work supported in part by  the Air Force Office of Scientific Research under Contract Nos. FA9550-10-1-0569,  by the Air Force Research Laboratories under contract FA8750-12-2-0186, by the National Science Foundation under Grant No. CIF-1217605, and by DARPA under Contract No. W911NF-12-1-0034.

\bibliographystyle{IEEEtran}
\bibliography{../references}
\balance

\end{document}